\documentclass{article}
\usepackage{hyperref}
\usepackage{amsmath} 
\usepackage{arxiv}
\usepackage{cite}
\usepackage{soul}
\usepackage{subcaption}
\usepackage[utf8]{inputenc} 
\usepackage[T1]{fontenc}    
\usepackage{hyperref}       
\usepackage{url}            
\usepackage{booktabs}       
\usepackage{amsfonts}       
\usepackage{nicefrac}       
\usepackage{microtype}      
\usepackage{lipsum}
\usepackage{graphicx}
 \usepackage{braket}
 
\title{Quantum Mechanics of  Particle on a torus knot:\\ Curvature and Torsion Effects}

\author{
  Dripto Biswas \\ 
  School of Physical Sciences\\
  National Institute of Science Education and Research\\
  Odisha, India \\
  \texttt{dripto.biswas@niser.ac.in}
   \\
   \And
  Subir Ghosh\\
  Physics and Applied Mathematics Unit \\
  Indian Statistical Institute, Kolkata\\
  West Bengal, India \\
  \texttt{subirghosh20@gmail.com}\\
}
\DeclareUnicodeCharacter{2212}{-}
\begin{document}
\maketitle

\begin{abstract}
Constraints play an important role in dynamical systems. However, the subtle effect of constraints in quantum mechanics is not very well studied. In the present work we concentrate on the quantum dynamics of a spin-less, point particle moving on a non-trivial torus knot. We explicitly take into account the role of curvature and torsion, generated by the constraints that keep the particle on the knot. We exploit the   "Geometry Induced Potential (GIP) approach" to construct the Schrodinger equation  for the dynamical system, obtaining thereby new results in terms of particle energy eigenvalues and eigenfunctions. We compare our results with  existing literature that completely ignored the contributions of curvature and torsion. In particular, we explicitly show how the "knottedness" of the path influences the results. In the process we have revealed a (possibly un-noticed) "topological invariant".
\end{abstract}


\section{Introduction}\label{sec:intro}

In recent times, theoretical development in quantum physics on curved low-dimensional
spaces has attracted a lot attention mainly due to the improved expertise  in synthesizing low-dimensional nanostructures with curved geometries \cite{schmidt,ahn,mei,ko,park}. The  approach of  De Witt \cite{de} was to directly consider the quantum particle motion in curved $n$-dimensional space.  On the other hand, the method suggested by  Jensen and Koppe \cite{koppe}
and by da Costa \cite{costa} was to treat the quantum dynamics in unconstrained $(n+1)$-dimensional
Euclidean space in the presence of appropriate  confining (or constraint) forces that restrict the particle motion on the designated surface. The important extension of \cite{costa} was performed by Wang et.al. \cite{Wang} (see also \cite{ortix}) who introduced the torsion term in the effective potential (experienced by the particle) in addition to the curvature term already incorporated by \cite{costa}.  Even though in classical mechanics the two procedures  are complimentary and equivalent, it turns out that the situation is different in quantum physics. In fact the curved space quantization programme of de Witt \cite{de} is plagued with  operator ordering ambiguities whereas the constrained dynamics framework of \cite{koppe,costa} (see also \cite{silva}) is well defined and  rigorous. Subsequent works in this area are: 
 a path integral formulation \cite{pi}, effects on the eigenstates of nanostructures \cite{ei,grave}, interaction with
an electromagnetic potential \cite{em,ferrari,oliv,silva2}, modeling of bound states on conical surfaces \cite{con,filg,Du}, spinorbit
interaction \cite{entin,gentile,ortix}, electronic transport in nanotubes \cite{santos,marchi}, and bent waveguides \cite{krej,peter,campo,haag}.

However, at present, there is a growing interest in studying the quantum mechanics  of 
 particles constrained to move on specified space curves. The analysis is directly applicable to complex three dimensional nanostructures such as  helical nanowires \cite{nanow} or multiple helices, toroids, and conical spirals \cite{nanoc}. The present paper deals with quantum motion of a spinless particle along a {\it{torus knot}}.
 
 The connection between the abstract mathematical  Knot theory and physics dates back to 
 the seminal work of Jones \cite{jo}, eg. Jones
polynomial (related to knot invariants) that were connected to topological quantum field theory by Witten \cite{wit1,wit2}. Via Temperley-Lieb
relations \cite{tem} direct connections between knot invaiants and statistical mechanical models (such as Q-state Potts model)
was established. Now knot theory has entered mathematical biology through the knotted structure  of certain DNA molecules. Knotted structures in physical fields appeared first in the
hypothesis of Lord Kelvin’s vortex atom model. The first concrete indication of knot-like structure
in physics appeared in the work of Faddeev and Niemi \cite{fad} who identified such structures with $(3+1)$-dimensional stable finite energy solitons and suggested that similar structures might be observed in nematic liquid crystals and $^3He$ superfluids. Only recently they were visible experimentally in diverse physical systems, such as in fluid vortex lines, singular lines of optical fields, and topological defects. Probably one of the simplest and
most relevant example of a knot is the torus knot. It is a knot that lies on the surface of an unknotted
torus in three dimensional space. A beautiful visualization of general (p, q)-torus knots can be found
in \cite{ke,new1} where a family of exact knotted solutions to Maxwell's equations in free space are represented
as torus knots and links constructed out of the electric and magnetic field lines that, surprisingly,
persist for all time.
Although this is not directly along the lines of our paper, it is still important to realize the
importance of toroidal geometry in physics and compactified quantum field theory on a torus (for a
review and exhaustive list of references see \cite{kh}). Indeed, it has become extremely relevant in recent
times because of the growing interest in finite size effects especially due to the vastly improved
experimental scenario. The significance of toroidal topology is that the local properties of the spacetime
remain unchanged and the non-trivial topological effect manifests, for example, in correlation
functions, which are non-local properties of the system.

It is now time to be more specific and concentrate on the problem at hand. In the present paper
we consider a different kind of influence of knot on physics where a particle is constrained to move
on a torus knot embedded in Euclidean three-space ($\mathbb{R}^3$). This type of generic toy model of particles
moving on a predetermined path (of various levels of complexity) actually helps in understanding
deeper issues, such as existence of inequivalent quantizations of a given classical system \cite{oh}, the role
of topology in the definition of the vacuum state in gauge theories \cite{col}, band structure of solids \cite{raj},
generalized spin and statistics of the anyonic type \cite{wil}, and the study of mathematically interesting
algebras of quantum observables on spaces with non-trivial topology [11], to name a few. Since a
particle moving on a circle is identified with a (single) rotor, a particle on a torus, considered here,
can be identified with a double-rotor, acting as a non-planar extension of the planar rotor \cite{sreedhar}. The simple problem of a particle on a ring, has been studied in detail for a long time. It has been used to explain the band structure of solids \cite{ohnuki} and to study the effects of non-trivial topological spaces on operator algebras \cite{flo}, among many others. A possible precursor to our problem is the work of \cite{taka}.

However, for a particle on a torus knot, the problem quickly becomes analytically complex. The primary difference being that knots form a class of non-contractible loops, due to the torus, being a surface of \textit{genus} 1. This makes the problem, interesting to investigate, and obtain the limits, in which the results for a particle on a ring can be reproduced. Quantum dynamics on torus knots can have varied applications, among which is the study the of topological insulators \cite{moore}. 

It has been shown in \cite{silva,costa}, that the quantum Hamiltonian for particles constrained to move on surfaces or curves in Euclidean space $\mathbb{R}^n$, contain a unique\cite{costa} Geometry-Induced Potential (GIP), which depends on the local property of curvature\cite{silva}. Furthermore, \cite{costa,silva}, also show that, this technique of introducing a confining potential, is free from operator ordering ambiguities, as encountered by earlier workers while trying to canonically quantize the classical Lagrangian for the problem. In the present work, we will  investigate the quantum dynamics of a particle constrained to move on a torus-knot, starting from the GIP approach.

The paper is organized as follows. In Section \ref{sec:potential}, we discuss construction of a generic GIP and show how the curvature and torsion terms appear in the effective potential. Section \ref{sec:torus}, provides a brief discussion of the torus knot along with our particular parameterization of the torus knot (Sub-section \ref{subsec:torcoor}), explicit expressions for curvature (Sub-section \ref{subsec:k}) and torsion (Sub-section \ref{subsec:t}) terms. In Sub-section \ref{subsec:torcur}, we show comparative strengths of the curvature and torsion terms in a numerical study. In Section \ref{sec:schro}, we write down the Schrodinger equation  and reveal the previously advertised "topological invariant" in question. In Section \ref{sec:energy}, we solve the Schrodinger equation in a thin-torus limit where analytic results are possible to derive. We also point out how the torus knot parameter appears in a non-trivial way. In Section \ref{sec:numerical}, we take recourse to a numerical simulation where we consider the full solution of the Schrodinger equation. We conclude in Section \ref{sec:conclusion}, with a summary of the present work along with open problems to be pursued in future.

\section{Generic form of effective potential from GIP}\label{sec:potential}

We consider a $d$-dimensional surface, $N^d$ embedded in a manifold, $\mathbb{M}^{d+k}$ of $d+k$ dimensions \cite{silva}. If we use the intrinsic coordinates of $N^d$, it does not invoke any property of $\mathbb{M}^{d+k}$. However, if we consider a constraining potential approach, in which the motion of the particle in constrained in $\mathbb{M}^{d+k}$, on $N^d$, due to some associated potential, it may also depend on properties of the ambient space $\mathbb{M}^{d+k}$\cite{silva}.

Jensen and Koppe(1971)\cite{koppe}, solved the problem of finding the equations of dynamics of a particle constrained to move on a surface, by starting from the equations for a particle in the region between two parallel surfaces, with homogeneous boundary conditions, and then taking the limit as the distance between them goes to zero.
Da Costa, also obtained the same equations, by considering a strong confining potential, corresponding to the desired surface or curve. 

In this section, we will briefly outline the derivation of da Costa \cite{costa} together with the modification by Wang et.al \cite{Wang}. Let us consider the constrained motion of a particle along a space curve $C$ given by $\vec r(q_1)$ with $q_1$ given by the arc-length. The close proximity of $C$ in $3$-space is denoted by 
\begin{equation}\label{1}
    \vec R(q_1,q_2,q_3)=\vec r(q_1) + q_2 \hat{n} + q_3 \hat{b},
\end{equation}
with $\hat{t} =\frac{\partial\vec r/\partial q_1}{||\partial\vec r/\partial q_1||} $ denoting the unit tangent vector and unit normal and binormal vectors given by, $\hat{n}$ and $\hat{b}$, respectively, along with 
\begin{equation}\label{2}
   \frac{\partial\vec t/\partial q_1}{||\partial\vec t/\partial q_1||}=\kappa \hat{n};~~ \frac{\partial\vec b/\partial q_1}{||\partial\vec b/\partial q_1||}=-\tau \hat{n},
\end{equation}
where $\kappa$ and $\tau$ are the curvature and torsion respectively. 

To restrict the particle motion along $C$, one may choose a binding potential $V_{\omega}(q_2,q_3)$(independent of $q_1$), where $\omega$ is a 'squeezing parameter' \cite{costa}, which decides the strength of the potential, such that,
\begin{equation}\label{potlim}
\lim_{\omega \rightarrow \infty} V_{\omega}(q_2,q_3) = \begin{cases}
0, \hspace{2mm} q_2= q_3=0\\
\infty,\hspace{2mm} q_2\ne 0, q_3\ne 0.\\
\end{cases}
\end{equation}
Following the approach of Wang et. al. \cite{Wang}, we note that the Hamiltonian in the plane, normal to $C$, may be written as,
\begin{equation}\label{HN}
\hat{H}_N = \hat{H}_n + \hat{H}_b,
\end{equation}
where, $\hat{H}_n$ and $\hat{H}_b$ are the normal and binormal components of $\hat{H}_N$ respectively, and are given by,
\begin{equation}\label{Hn}
    \hat{H}_n = -\frac{\hbar^2}{2m}\partial_2^2 + \frac{1}{2}m\omega q_2^2,~~
\hat{H}_b = -\frac{\hbar^2}{2m}\partial_3^2 + \frac{1}{2}m\omega q_3^2.
\end{equation}
 $\omega$ is the squeezing parameter considered, in the potential $V_{\omega}(q_2,q_3)$ (\ref{potlim}).
Solving the 1-D Harmonic potential problem  we get \cite{Wang} ,
\begin{equation}\label{Xn}
\ket{\chi_{0n}} = \nu^{\frac{1}{2}}\pi^{-\frac{1}{4}}e^{\frac{(\nu q_2)^2}{2}},~~
\ket{\chi_{0b}} = \nu^{\frac{1}{2}}\pi^{-\frac{1}{4}}e^{\frac{(\nu q_3)^2}{2}},
\end{equation}
where $\ket{\chi_{0n}}$ and $\ket{\chi_{0b}}$ are the ground state wavefunctions in the normal and binormal directions and $\nu=\sqrt{m\omega /\hbar}$. For very large $\omega$ the motion in the normal plane gets frozen leaving only the motion along the space curve.

The quantum Hamiltonian and the corresponding Schrodinger equation  for such a curve is obtained in \cite{costa},
\begin{equation}\label{genschro}
\frac{-\hbar^2}{2m}\left[\frac{1}{(1- \kappa f)}\frac{\partial }{\partial q_1}\left(\frac{1}{(1- \kappa f)}\frac{\partial \psi}{\partial q_1}\right) + \sum_{j=2}^3\left(\frac{\partial^2 \psi}{\partial q_j^2} \frac{\partial}{\partial q_j} \ln(1 - \kappa f)\frac{\partial \psi}{\partial q_j}\right)\right] + V_{\omega}(q_2, q_3)\psi = i\hbar\frac{\partial \psi}{\partial t},
\end{equation}
where, $\kappa = \kappa(q_1)$, is the curvature of $\vec r(q_1)$ at $q_1$, and $f(q_1) = q_2\cos(\theta) + q_3\sin(\theta)$. 

A more convenient form of  wavefunction is given by,
\[\chi(q_1,q_2,q_3) = \sqrt{(1-\kappa f)}\psi,\]
and substituting back in (\ref{genschro}), we obtain,

\begin{equation}\label{genschro2}
-\frac{\hbar^2}{2m}\frac{1}{\sqrt{(1-\kappa f)}}\frac{\partial}{\partial q_1}\left(\frac{1}{(1-\kappa f)}\frac{\partial}{\partial q_1}\frac{\chi}{\sqrt{(1- \kappa f)}}\right) -\frac{\hbar^2}{8m}\frac{\kappa^2}{(1 - \kappa f)^2}\chi -\frac{\hbar^2}{2m}\left(\frac{\partial^2 \chi}{\partial q_2^2} + \frac{\partial^2 \chi}{\partial q_3^2}\right) + V_{\omega}(q_2, q_3)\chi = i\hbar\frac{\partial \chi}{\partial t}.
\end{equation}

Now, from the limiting behaviour of $V_{\omega}(q_2,q_3)$ in (\ref{potlim}), we can put $f \rightarrow 0$ in (\ref{genschro2}) to obtain \cite{costa},

\begin{equation}\label{endschro1}
-\frac{\hbar^2}{2m}\frac{\partial^2 \chi}{\partial q_1^2} -\frac{\hbar^2}{8m}\kappa(q_1)^2\chi -\frac{\hbar^2}{2m}\left(\frac{\partial^2 \chi}{\partial q_2^2} + \frac{\partial^2 \chi}{\partial q_3^2}\right) + V_{\omega}(q_2, q_3)\chi = i\hbar\frac{\partial \chi}{\partial t}.
\end{equation}
(\ref{endschro1}), is separated, by setting $\chi = \chi_1(q_1,t)\chi_2(q_2,q_3,t)$, whereby we get,
for, $\chi_1$,
\begin{equation}\label{endschro2}
-\frac{\hbar^2}{2m}\frac{\partial^2 \chi_1}{\partial q_1^2}-\frac{\hbar^2}{8m}\kappa(q_1)^2\chi_1 = i\hbar\frac{\partial \chi_1}{\partial t}.
\end{equation}

Therefore, the modified Schrodinger equation for a curve, which can be read off (\ref{endschro2}) is, \cite{costa,silva},

\begin{equation}\label{schro}
    i\hbar\frac{\partial \Psi}{\partial t} = \frac{-\hbar^2}{2m}\left(\Delta_s + \frac{\kappa^2}{4}\right)\Psi,
\end{equation}
where, $\Delta_s = \frac{d^2}{ds^2}$ is the Laplace-Beltrami operator, or in this case, just the Laplacian on the curve, with respect to the arc-length coordinate $s$, and $\kappa$ is the local curvature of the path $C$. Here, $\Psi = \chi_1$ is the wavefunction on the curve.

However this is not the end of the story since as Wang et. al. \cite{Wang} pointed out, 
(\ref{schro}) correctly describes the quantum mechanics for space curves with $\tau = 0$, but fails to do so for curves with non-zero torsion. The proper way \cite{Wang} is to relax the condition (\ref{potlim}) to
\begin{equation}\label{wan}
\lim_{\omega \rightarrow \infty} V_{\omega}(q_2,q_3) = \begin{cases}
0, \hspace{3.6mm} \mid q_2\mid \leq \epsilon_2, \mid q_3\mid \leq \epsilon_3\\
\infty,\hspace{2mm} \mid q_2\mid >  \epsilon_2,~\mid q_3\mid > \epsilon_3.\\
\end{cases}
\end{equation}
where $\epsilon_2,\epsilon_3$ are length scales in the plane normal to $C$. It can be shown  that the effective Hamiltonian is then given by \cite{ortix,Wang}
\begin{equation}\label{Heff}
\hat{H}_{eff} = \lim_{\omega \rightarrow 0}\bra{\chi_{0n,0b}}F^{-\frac{1}{2}}\hat{H}_v F^{\frac{1}{2}} - \hat{H}_N\ket{\chi_{0n,0b}}
\end{equation}
where, $\hat{H}_v = -\frac{\hbar^2}{2m}\nabla^2 + V_{\omega}(q_2,q_3)$, $F = (1 - \kappa q_2)$ and $\ket{\chi_{0n,0b}} = \ket{\chi_{0n}} \otimes \ket{\chi_{0b}}$.{\footnote{Wang et. al. \cite{Wang} also notes that in \cite{costa} the effective Hamiltonian was taken as $\hat{H}_{eff} = \lim_{\omega \rightarrow 0}\bra{\chi_{0n,0b}}F^{-\frac{1}{2}}\hat{H}_v F^{\frac{1}{2}} \ket{\chi_{0n,0b}}- \hat{H}_N$ and thereby missed the torsion contribution.}}

Thus the final form of the effective hamiltonian is given by \cite{Wang},
\begin{equation}\label{Hfinal}
    \hat{H}_{eff} = -\frac{\hbar^2}{2m}\left(\nabla^2_s + \frac{\kappa^2}{4} - \frac{\tau^2}{2} \right) 
\end{equation}
where $\frac{\tau^2}{2}$ is the additional torsion-induced potential. 

We emphasize that we will be working with the effective hamiltonian given in (\ref{Hfinal}) which is very different qualitatively from the one considered recently in the work \cite{sreedhar}. For the torus knot in question both curvature and torsion contributions are appreciable. We will consider situations where $\frac{\tau^2}{2} << \frac{\kappa^2}{4}$ so that one can ignore the torsion effect. Furthermore, in this limit, we will reveal a curious expression in the effective hamiltonian that we have advertised as a "topological invariant". In the next section we will introduce the torus knot and re-express the effective hamiltonian (\ref{Hfinal}) in toroidal coordinates that respects the natural symmetry of the regular torus.

\section{Torus Knots - Definitions and Properties}\label{sec:torus}
A torus  is a surface of revolution generated by revolving a circle in three-dimensional space about an axis that is coplanar with the circle. If the axis of revolution does not touch the circle, the surface has a ring shape  and is called a torus of revolution.{\footnote{As the distance from the axis of revolution decreases, the ring torus becomes a horn torus, then a spindle torus, and finally degenerates into a sphere. }} Torus knots are a special class of non-self-intersecting, closed curves, wound on the surface of a geometric torus in $\mathbb{R}^3$.  The family of knots appear because 
the first homotopy group for the torus  is $\pi 1(torus) \approx Z_1\times Z_1$. 
\begin{figure}[!htbp] 
  \centering
  \includegraphics{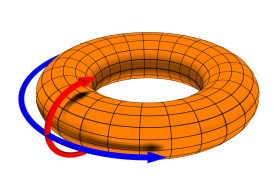}
  \includegraphics[scale=0.7]{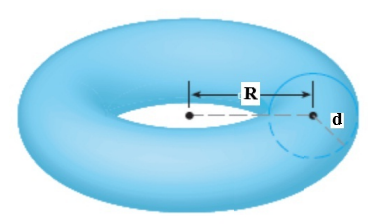}
  \caption{(Left) Red arrow depicts poloidal direction, while blue arrow depicts toroidal direction. (Right) The major and minor radii of a torus (Picture Courtesy - P. Das et. al.\cite{sg}) . \label{fig1}}
\end{figure}

\begin{figure}
\hspace{9mm}
\includegraphics[scale=0.9]{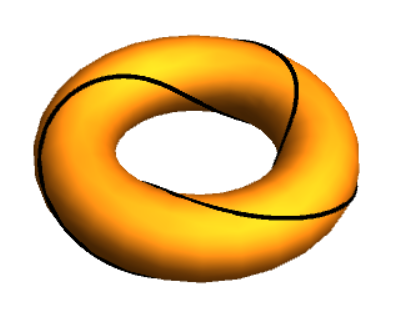}
\includegraphics[scale=0.9]{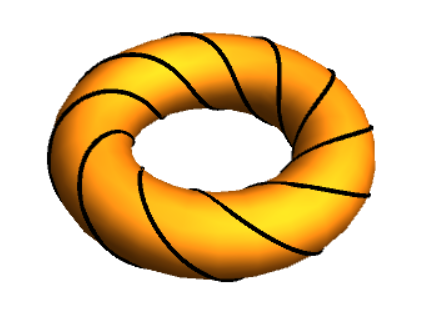}
\caption{(Left) A trefoil knot, or a $(2,3)-$torus knot, is the simplest non-trivial knot. (Right) A (3,10)-torus knot.} \label{fig2}
    \label{fig:my_label}
\end{figure}

For our purposes, each knot is defined using 2 co-prime integers, $p,q \in \mathbb{N}$.
A $(p,q)-$torus knot winds $p$ times in the toroidal direction, about the axis of rotational symmetry of the torus and $q$ times in the poloidal direction, about the cross-sectional axis of symmetry (see Fig.\ref{fig1}). The quantity $\alpha = \frac{q}{p}$ is called the 'winding number' of the $(p,q)-$ torus knot, and is a simple measure of the complexity of the knot. One may note immediately, that two different torus knots (differing in at least one of $p$ or $q$), will have different values of $\alpha$. In this sense, $\alpha$, may be regarded as the 'unique' identity of a $(p,q)-$torus knot. Some popular examples of torus  knots are depicted in Fig. \ref{fig2}, below.

\subsection{Parametrisation of torus knots in toroidal coordinates}\label{subsec:torcoor}
Henceforth, we shall use toroidal coordinates, to exploit the inherent symmetry of the problem. We consider a torus of major radius $R$ and minor radius $d$ (see Fig. \ref{fig1} above). Then, we can define the parameter $a^2 = R^2 - d^2$, and the \textit{aspect ratio} of the torus, $\cosh(\eta_0) = \frac{R}{d}$. For a thin torus, $a>>0$, the \textit{aspect ratio} can be quite large.

The parametrisation for torus knots in toroidal coordinates, is then given by\cite{sreedhar},
\begin{equation}\label{toroipara}
    x= \frac{a \sinh\eta_0 \cos\phi}{\cosh\eta_0 - \cos(\alpha \phi)},~ y = \frac{a \sinh\eta_0 \sin\phi}{\cosh\eta_0 - \cos(\alpha \phi)},~z = \frac{-a \sin(\alpha\phi)}{\cosh\eta_0 - \cos(\alpha \phi)}.
\end{equation}
where $x,y,z$ are Cartesian coordinates, $\eta_0$ fixes the toroidal surface on which the knot is wound, $\alpha$ is the winding number of the knot as defined above, and $0<\phi<2\pi p$. Note that the winding number condition has been  imposed so that the only variable is $\phi$. 

Since the relations in Section 2 involved the arc length $s$ we need to express $s$ in terms of $\phi $. Computing the differential arc-length $ds$, from (\ref{toroipara}), we get,
\[ds^2 = dx^2 + dy^2 + dz^2 = a^2\sigma^4(\lambda^2 - 1)d\phi^2,\]
from which we obtain 
\begin{equation}\label{dphids}
\frac{d\phi}{ds} = \frac{1}{a  \sigma^2\sqrt{\lambda^2 -1}} = \frac{1}{a\beta \sigma^2}.
\end{equation}
where  $\sigma = \frac{1}{\sqrt{b - \cos(\alpha \phi)}}$, $\lambda^2 = \alpha^2 + c^2$, $c = \sinh(\eta_0)$, $b = \cosh(\eta_0)$, and $\beta^2 = \lambda^2 - 1$.
As we will see in the next section, (\ref{dphids}) and it's derivative will be required to rewrite the Schrodinger equation (\ref{Hfinal}).

\subsection{Curvature of a torus-knot}\label{subsec:k}
Let us start by rewriting the space curve $C$ of previous section as $\vec \gamma(\phi) = x(\phi)\hat{i} + y(\phi)\hat{j} + z(\phi)\hat{k}$. The curvature of a general curve $\vec \gamma(s)$, in Euclidean space, is given by,

\begin{equation}\label{k_theo}
\kappa(s) = \left|\frac{d^2 \vec \gamma(s)}{ds^2}\right|,
\end{equation}
where $s$ is the arc-length parameter.

For a general parametrisation using $\phi$, given by $\vec \gamma(\phi)$, (\ref{k_theo}) is rewritten as,
\begin{equation}\label{k_xyz}
\kappa = \frac{\sqrt{(z''y' - y''z')^2 + (x''z' - z''x')^2 + (y''x' - x''y')^2}}{(x'^2 + y'^2 + z'^2)^\frac{3}{2}}.
\end{equation}
Using (\ref{dphids}) and (\ref{k_xyz}), we get, 
\begin{equation}\label{k_val}
\frac{\kappa^2}{4} = \frac{\left[2 - 6\alpha^2 + 4\alpha^4 + 2\cos(2\alpha\phi) - 2\alpha^2\cos(2\alpha\phi) + 8(\alpha^2 - 1)b\cos(\alpha\phi) + 4b^2\right]c^2}{4 a^2\left[2b^2 + 2\alpha^2 - 2\right]^2},
\end{equation}
where, all symbols are as defined above.

\subsection{Torsion of a torus-knot}\label{subsec:t}
For a curve $\vec \gamma(s)$, the unit normal $\hat n$ and binormal $\hat b$ are given by,
\[\hat{n} = \frac{1}{\kappa}\frac{d\hat{t}}{ds}, \hat{b} = \hat{t}\times\hat{n},\]
respectively.
Then, the torsion is simply given by,
\begin{equation}\label{tor(s)}
\tau(s) = -\hat{n}(s)\cdot\frac{d\hat{b}}{ds}
\end{equation}
Similar to the curvature, the torsion $\tau(\phi)$, of a curve parametrized by $\phi$,  is given by,
\begin{equation}\label{tau}
\tau = \frac{x'''(y'z'' - y''z') + y'''(x''z' - x'z'') + z'''(x'y'' - x''y')}{(y'z'' - y''z')^2 +(x''z' - x'z'')^2 + (x'y'' - x''y')^2 }.
\end{equation}
Substituting in (\ref{tau}), $x(\phi),y(\phi),z(\phi)$ from (\ref{toroipara}), and calculating $\frac{\tau^2}{2}$ we get,
\begin{equation}\label{tauval}
\frac{\tau^2}{2} = \frac{32\alpha^2(a^2-1)^2(\cos(\alpha\phi) - b)^4(b\cos(\alpha\phi) + \alpha^2 - 1)^2}{a^2(2b^2 + 2\alpha^2 - 2)^2\left[3 - 6\alpha^2 + 4\alpha^4 -2(\alpha^2 - 1)\cos(2\alpha\phi) + 8b(\alpha^2 - 1)\cos(\alpha\phi) + 4b^2 - 1\right]^2}.
\end{equation}
In the following sub-section, we show numerically, using Wolfram Mathematica, the regions in parameter space of the torus knot where torsion effect dominates over curvature effects and where the reverse occurs. Quite interestingly, through $\alpha$ and $b$ parameters, properties of both the knot and the host torus get intertwined. We perform this exercise because   later on we will study analytic solutions for energy eigenvalues and eigenfunctions that are difficult to obtain if the torsion term is present in the effective hamiltonian. The numerical study will help us to concentrate on those sectors where the torsion contribution can be neglected, thereby reducing (\ref{Hfinal}) into (\ref{schro}).

\subsection{Numerical comparison of torsion and curvature terms}\label{subsec:torcur}
\begin{figure}
\hspace{-18mm}
\includegraphics[scale=0.73]{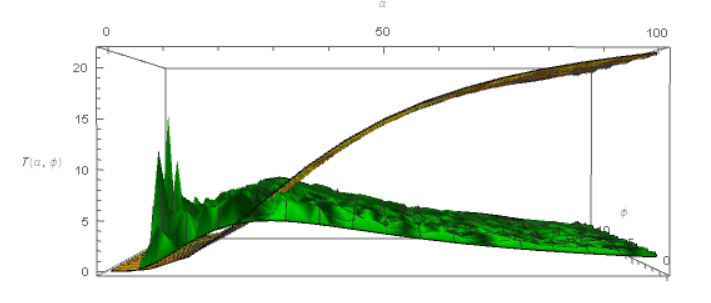}
\includegraphics[scale=0.73]{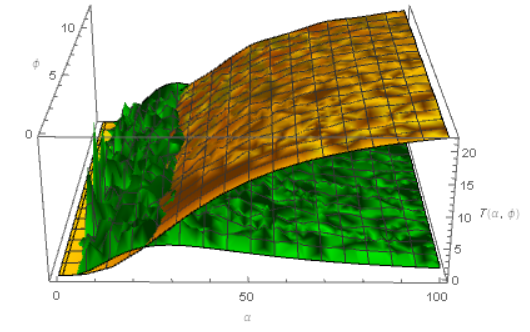}
\caption{(Left) The green surface represents $\frac{\tau(\alpha,\phi)^2}{2}$, while the yellow surface depicts $\frac{\kappa(\alpha,\phi)^2}{4}$. (Right) The same plot, at a slightly different viewing-angle reveals the 'low $\alpha$' region, where torsion dominates curvature. The opposite happens for large $\alpha$ with $\alpha\approx b$ providing roughly the demarcation line.} \label{fig3}
    \label{fig:torcur}
\end{figure}

We use Mathematica, to draw surface plots of $\frac{\tau(\alpha,\phi)^2}{2}$ and $\frac{\kappa(\alpha,\phi)^2}{4}$, by varying both $\alpha$ and 
$\phi$, for $0\le \phi \le 2p\pi$ and $1\le \alpha \le 100$. We  consider essentially a thin-torus with $R=4$ and $d=1$ ($\eta_0 = 4$) to be the torus parameters for this numerical analysis. The plots obtained are depicted in Fig. \ref{fig:torcur}.
The vertical axis, denoted by $T(\alpha,\phi)$, is the common axis for both $\frac{\tau(\alpha,\phi)^2}{2}$ and $\frac{\kappa(\alpha,\phi)^2}{4}$.

It can be seen from Figure. \ref{fig:torcur}, that $\frac{\kappa^2}{4}$ begins to dominate $\frac{\tau^2}{2}$, for $\alpha \approx 25 \approx \cosh(\eta_0)$. Therefore, for $\mathcal{O}(\alpha^2) \ge \mathcal{O}(b^2)$, the torsion term in (\ref{Hfinal}), is much smaller compared to the curvature term.

\begin{figure}
\includegraphics[scale=0.8]{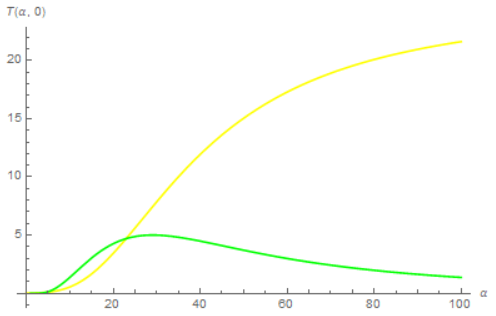}
\includegraphics[scale=0.8]{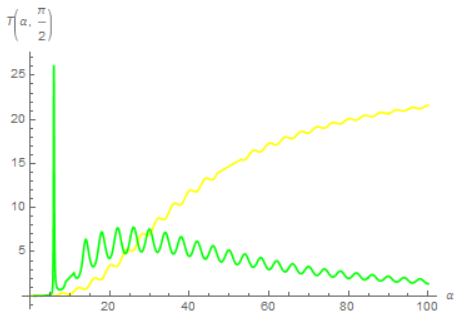}
\caption{(Left) The green curve represents an angular slice of the torsion surface plot in Fig. \ref{fig:torcur}, corresponding to $\phi=0$, while the yellow surface depicts the same angular slice of the curvature surface plot in Fig. \ref{fig:torcur}. (Right) The same plot, for $\phi=\frac{\pi}{2}.$} \label{fig:4}
    \label{fig:torcur2d}
\end{figure}

Fig. \ref{fig:torcur2d} shows two projections of the surface plots for fixed $\phi$, for better clarity, clearly showing how $\frac{\kappa^2(\alpha,\phi)}{4}$ begins to dominate $\frac{\tau^2(\alpha,\phi)}{2}$, for large $\alpha$. Intuitively it is not difficult to understand the reason: for large $\alpha=q/p$, the   number of poloidal loops increase and for a thin torus the poloidal loops are much more curved than the toroidal loops.

Hence we will be considering from now on large $\alpha$ torus knots on a thin-torus so that the torsion term can be safely neglected. It is important to note that we keep the curvature contribution which will yield new features in the expression of particle wave function. We will compare and contrast with the result obtained by  Sreedhar \cite{sreedhar} who did not take in to account the curvature term (or the torsion term).

\section{ Schrodinger Equation for toroidal coordinates for particle on torus loop}\label{sec:schro}

We now proceed to solve the time-independent Schrodinger equation for (\ref{schro}) (with torsion term neglected).

The Laplacian $\frac{d^2}{ds^2}$, can be reparametrised in terms of $\phi$, whereby, we obtain the time-independent Schrodinger Equation, in toroidal coordinates, to be,
\begin{equation}\label{schro2}
\left[\left(\frac{d\phi}{ds}\right)^2\frac{d^2}{d\phi^2} + \left(\frac{d^2\phi}{ds^2}\right)\frac{d}{d\phi} + \frac{\kappa^2}{4}\right]\psi = -\epsilon \psi,
\end{equation}
where, $\epsilon = \frac{2m E}{\hbar^2}$, and $E$ is the energy.

We now consider $\psi = f(\phi)G(\phi)$, and substitute it back in (\ref{schro2}). Putting the coefficient of $G'$ to be zero, in this case, we obtain,

\begin{equation}\label{sigmaschro}
\left[\frac{d^2}{d\phi^2} + \left(\frac{\sigma\sigma'' - 2\sigma'^2}{\sigma^2}\right) + (a\beta)^2\sigma^4\left(\frac{\kappa^2(\phi)}{4} + \epsilon\right)\right]G = 0,
\end{equation}
where we also get, \[f(\phi) = \sigma(\phi) = \frac{1}{\sqrt{b - \cos(\alpha\phi)}}.\]
For $\kappa (\phi)=0$, (\ref{sigmaschro}) reduces to the one considered in \cite{sreedhar}. However, {\it a priori} there is no reason to drop the curvature term. Clearly it will affect the results in a non-trivial way since $\kappa(\phi)$ is not a numerical constant. Here, we clearly see the distinction between canonical quantization, and confining potential approaches to the same problem, as mentioned in \cite{costa}. 

Now comes the unexpected result. Substituting, the expression for $\sigma$, in (\ref{sigmaschro}), we obtain,
\begin{equation}\label{sigma}
\left(\frac{\sigma\sigma'' - 2\sigma'^2}{\sigma^2}\right) =  \frac{\alpha^2\left[3 - 4b\cos(\alpha \phi) + \cos(2\alpha \phi)  \right]}{8\left[b - \cos(\alpha \phi)\right]^2}.
\end{equation}
Notice that the $\phi-$dependent terms, other than the energy, in (\ref{sigmaschro}), add up to a {\it numerical constant}, consisting of the winding number of the torus loop and aspect ratio of the torus. We can see this, by adding $\left(a^2\beta^2\sigma^4\frac{\kappa^2}{4}\right)$ and (\ref{sigma}) to get,
\begin{equation}\label{const}
   \left(\frac{\sigma\sigma'' - 2\sigma'^2}{\sigma^2}\right) + a^2\beta^2\sigma^4\frac{\kappa^2}{4} = \frac{b^2 + \alpha^4 - 1}{4(b^2 + \alpha^2 - 1)}. 
\end{equation}
We have referred to this number 
\begin{equation}\label{const1}
 \Gamma = \frac{b^2 + \alpha^4 - 1}{4(b^2 + \alpha^2 - 1)}
 \end{equation}
 as a "topological invariant". The structure of $\Gamma$, is  interesting, since it incorporates the main property of the knot, through the parameter $\alpha$, as well as the property of the torus on which it is wound, through the parameter $b$. This is one of our intriguing results.
 
Substituting back in (\ref{sigmaschro}), we get,
\begin{equation}\label{constschro}
\left[\frac{d^2}{d\phi^2} + \frac{b^2 + \alpha^4 - 1}{4(b^2 + \alpha^2 - 1)} + (a\beta)^2\sigma^4\epsilon\right]G = 0.
\end{equation}

Thereby, we shall solve (\ref{constschro}) for $G(\phi)$. The complete solutions to (\ref{schro}) may  be obtained by multiplying $G$ with $\sigma(\phi)$. We shall also obtain the energy eigenvalues for the thin-torus approximation, by imposing the condition of periodicity on the wavefunctions.

\section{Solving for the energy eigenvalues}\label{sec:energy}

To have a better understanding of (\ref{constschro}), we expand the third term in the equation, as a binomial series, to obtain,
\begin{equation}\label{hillschro}
   \left[\frac{d^2}{d\phi^2} + \left(\Gamma + \frac{a^2\beta^2\epsilon}{b^2}\sum_{k=0}^\infty(-1)^k \binom{k+1}{k}\left(\frac{\cos(\alpha\phi)}{b}\right)^k  \right)\right]G = 0.
\end{equation}

Note that, this expansion is meaningful, since $b = \cosh(\eta_0) > 1 \ge \cos(\alpha\phi)$, $\forall \eta_0 \in (0,\infty)$. 

For thin-torus approximation, we consider large $b$, for which the higher order terms in (\ref{hillschro}) may be neglected.

On substituting, $2z = \alpha\phi$, (\ref{hillschro}) can immediately be seen to be of the form,

\begin{equation}\label{hill}
\left[\frac{d^2}{dz^2} + \Theta_0 + \sum_{r=1}^\infty \Theta_{2r} \cos(2 r z)\right]G = 0,
\end{equation}
where the $\Theta_i$ are constants depending on the energy and torus knot parameters. (\ref{hill}) is the well-known Hill Equation. 

Although the Hill Equation was also obtained in \cite{sreedhar}, the parameters $\Theta_i$ we have obtained, differ from their results, since we have introduced the curvature term in the 'confining potential approach', unlike the canonical quantization used by them.

Since there are no closed-form, analytic solutions to (\ref{hill}), we look for solutions to simpler cases, arising in the thin-torus limit.

\subsection{Thin-torus approximation}\label{subsec:thintor}

For a thin-torus, we may assume $b^2 \approx c^2$, and neglect terms in (\ref{hillschro}), having overall order $\mathcal{O}\left(\frac{1}{b^2}\right)$ and above, in $\frac{1}{b}$. With these approximations, we obtain from (\ref{hillschro}),

\begin{equation}\label{mathieu}
   \left[\frac{d^2}{dz^2} + \left(\frac{4\Gamma}{\alpha^2} + \frac{4a^2\beta^2\epsilon}{b^2\alpha^2}\right) - 2\frac{a^2\beta^2\epsilon}{b^3}\cos(2z) \right]G = 0.
\end{equation}

This is the Mathieu differential equation, also obtained for the thin-torus limit, in \cite{sreedhar}, albeit with different coefficients. The applicable solutions to (\ref{mathieu}), with the correct boundary conditions, are given by Mathieu functions of fractional order $\nu = \frac{2n}{q}$. Note that, such a choice of $\nu$, implies that $z$ can have $2\pi q$ periodicity, which translates into the required $2\pi p$ periodicity of the wavefunctions $G(\phi)$ \cite{sreedhar,mathieu1}.

Considering only upto overall order $\mathcal{O}\left(\frac{1}{b}\right)$, the condition for periodic wavefunctions implies that \cite{ince},

\begin{equation}\label{energy1}
\Theta_0 = \nu^2 + \mathcal{O}\left(\frac{1}{b^2}\right)  = \frac{4n^2}{q^2}.
\end{equation}
We also have  \[\Theta_0 = \left(\frac{4\Gamma}{\alpha^2} + \frac{4a^2\beta^2\epsilon}{b^2\alpha^2}\right).\]
Solving for the energy $E_n$ from (\ref{energy1}), we get,
\[E_n = \left(\frac{n^2 b^2 \alpha^2}{a^2 \beta^2 q^2} - \frac{b^2 \Gamma}{a^2\beta^2}\right)\frac{\hbar^2}{2m}.\]
The second term in the parenthesis above, may be ignored, since it is just a constant, and we can shift our zero-reference for measuring the energy of the system, to obtain explicitly,
\begin{equation}\label{energyfin}
E_n = \frac{\hbar^2 n^2 b^2 }{2m a^2 p^2(\alpha^2 + c^2 - 1)} = \frac{\hbar^2n^2\cosh^2(\eta_0)}{2m a^2 p^2 (\alpha^2 + \sinh^2(\eta_0) - 1)}.
\end{equation}
(\ref{energyfin}), is our primary result. It is important to note that the above result  differs from the energy eigenvalues obtained in \cite{sreedhar} in a non-trivial way. Let us express our result (\ref{energyfin}) as,
\begin{equation}\label{e0}
E_n = E_{0,n}F(\eta,\alpha),
\end{equation}
where, 
\begin{equation}\label{e1}
E_{0,n} = \frac{n^2\hbar^2}{2ma^2p^2},~~
F(\eta,\alpha) = \frac{\cosh^2(\eta)}{(\alpha^{2} + \sinh^2(\eta) - 1)}.
\end{equation}
Here  $E_{0,n}$ refers to the energy eigenvalues obtained in \cite{sreedhar} and $F(\eta,\alpha)$ is the correction factor revealed in our work. Notice that $E_{0,n}$ depends only on $p$, which only partly characterizes the torus knot whereas the correction factor depends on $\alpha=q/p$ that fully specifies the knot. Intuitively one can understand this in the following way. For a thin torus, $p$ becomes the dominant feature of the knot (at least as far as energy is concerned) since it describes the rotation in the toroidal direction that is longer for the thin torus limit. For thin torus effect of rotation in the poloidal direction having shorter radius is comparatively small (again for the energy at least). Quite interestingly for sufficiently large $\alpha$ such that in the denominator of $F(\eta,\alpha)$, $(\alpha^{2} + \sinh^2(\eta) - 1)\approx \alpha^2$, the energy values reduce to
$$E_n=\frac{\hbar^2n^2\cosh^2(\eta_0)}{2m a^2 q^2 }$$
so that $q$, the number of poloidal loops, now takes over the role of $p$. In the extreme limit of a very thin torus, $F\approx 1$ such that the result for motion along a hoop is recovered. Clearly, the curvature term that we have considered has given rise to these qualitatively new features in the particle observable.
 
The general solutions to (\ref{mathieu}), are given by the Mathieu functions of fractional order $\nu$ \cite{mathieu2} which yields
\begin{equation}\label{gensol}
G = A \textit{se}_{\nu}(z,b) + B \textit{ce}_{\nu}(z,b),
\end{equation}
where, $A$ and $B$ may be determined from normalization conditions. The functions, $\textit{se}_{\nu}(z,b)$ and $\textit{ce}_{\nu}(z,b)$, are of the form \cite{ince},

\begin{equation}\label{se}
\textit{se}_{\nu}(z,b) = \sin(\nu z) + \frac{\hbar^2}{16 m b p^2}\sin(3\nu z) + \mathcal{O}\left(\frac{1}{b^2}\right),
\end{equation}
\begin{equation}\label{ce}
\textit{ce}_{\nu}(z,b) = \cos(\nu z) + \frac{\hbar^2}{16 m b p^2}\cos(3\nu z) + \mathcal{O}\left(\frac{1}{b^2}\right).
\end{equation}

\subsection{Brief discussion on the results obtained}\label{subsec:results}

Our result in (\ref{energyfin}), for a thin-torus, describes how the energy eigenvalues depends both on the geometric properties of the torus, through $b$ and $c$, as well as on the property of the torus-knot, via the 'unique' quantity $\alpha$ (Section \hyperref[sec:torus]{3}).

We notice that, for the case, $\mathcal{O}(\alpha^2) \approx \mathcal{O}(c^2)\approx \mathcal{O}(b^2)$, i.e. for large winding numbers $\alpha$, the obtained result deviates from that obtained in \cite{sreedhar}, even in the thin-torus limit. 

This is a crucial difference from existing literature(to the best of our knowledge), as the topological complexity of knots, manifested through the parameter $\alpha$ in (\ref{energyfin}), does indeed play a role in modifying the energy eigenvalues, even in the thin-torus limit, as opposed to the results obtained in Sreedhar \cite{sreedhar}, which remains unchanged for arbitrary $\alpha$ (with fixed values of $p$).

Note that the energy eigenvalues in (\ref{energyfin}),
degenerate into the well-known energy expression for a particle on a ring (with $\tau = 0$), if we consider $\mathcal{O}(b^2) \approx \mathcal{O}(c^2)$ and $\alpha^2 << c^2 \approx b^2$, i.e. for small winding numbers. 

We also observe from (\ref{se}) and (\ref{ce}), that for large values of $\mathcal{O}(b^2)$, the solutions degenerate into sine and cosine functions, which is expected for a particle on a ring (thin-torus knot) since a thin-torus knot, is essentially a putative circle.

The complete eigenfunctions of the original problem, may be obtained from (\ref{gensol}), by multiplying by $\sigma(\phi)$.
\subsection{Surface plot of the correction factor}\label{subsec:corrfactor}

\begin{figure}
\hspace{-15mm}
\includegraphics[scale=0.8]{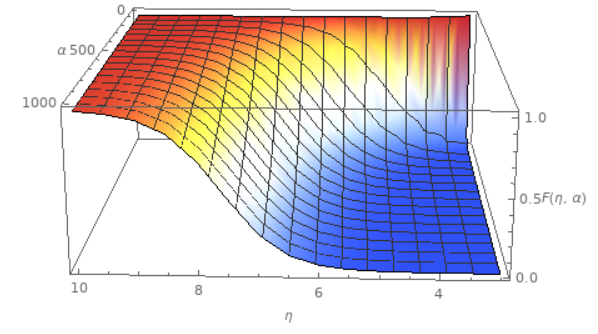}
\includegraphics[scale=0.8]{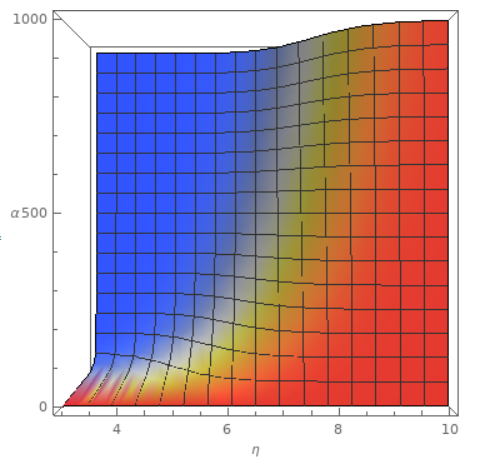}
\caption{(Left) Front-View of surface plot, with $\alpha \in (0,1000)$ and $\eta \in (3,10)$. (Right) Top-View of the same plot. In the red zone $F\approx 1$ where our results roughly agree with \cite{sreedhar} whereas in the blue region the results greatly disagree.} \label{fig3}
    \label{fig:my_label}
\end{figure}

In this subsection, we construct a graphical representation (Fig. \ref{fig3}) of the regions in $(\alpha,\eta)$ parameter space, where we indicate how  our results  deviate from \cite{sreedhar} resulting from the curvature effect. From the surface plot for $F(\eta,\alpha)$, in Fig. \ref{fig3}, we see that the factor $F(\eta,\alpha)\approx 1$, for low values of $\alpha$ and high values of $\eta$. Hence, for the parameters in the 'red' region, our result for the energy eigenvalue expression (\ref{energyfin}), reduces to the energy expression $E_{0,n}$, obtained by Sreedhar.

The 'bluer' regions depict subspaces (of the parameter space), with $\mathcal{O}(b^2) \approx \mathcal{O}(\alpha^2)$, while the 'reddish' region is for $\mathcal{O}(\alpha^2) << \mathcal{O}(b^2)$. This is in accordance with the predictions from the theoretical results, as discussed in Sub-section-\ref{subsec:results}.  However, our result deviates significantly from \cite{sreedhar} in the 'blue' regions of the surface plot, since we have correctly taken into account the curvature term.

\section{Numerical simulation }\label{sec:numerical}
Finally we briefly demonstrate the validity of the analytic expression of energy eigenvalues computed from the thin torus approximated expression, with a numerical result obtained by considering the exact equation (\ref{constschro}). We use Wolfram Mathematica, to numerically solve (\ref{constschro}), by moving $\epsilon$ to the R.H.S and multiplying by, $\frac{1}{\sigma^4}$ throughout, to obtain,

\begin{equation}\label{numschro}
(b - \cos(\alpha\phi))^2\left[\frac{d^2}{d\phi^2} + \Gamma\right]G(\phi) = -(a \beta)^2\epsilon G(\phi).
\end{equation}

The idea is to demonstrate and compare our thin-torus result with the numerical values to confirm the theoretical calculations, using 6 eigenvalues of this equation, for 2 cases of $\alpha=1.5$ and $\alpha=3.5$. For both cases, we chose $R=3$ and $d=1$ ($\eta_0=3$).

We solved (\ref{numschro}), subject to the boundary conditions,
\begin{equation}\label{bound1}
G(0) = G(2\pi p),
\end{equation}
and,
\begin{equation}\label{bound2}
G'(0) = G'(2\pi p)
\end{equation}
These two boundary conditions, also specifies the eigenvalues, since condition (\ref{bound2}), is true only for specific values of $\epsilon$, in (\ref{numschro}).

\begin{figure}
\hspace{-18mm}\includegraphics[scale=0.78]{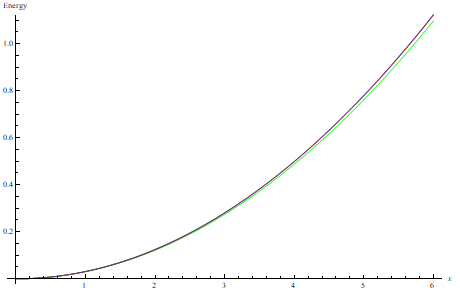}
\includegraphics[scale=0.78]{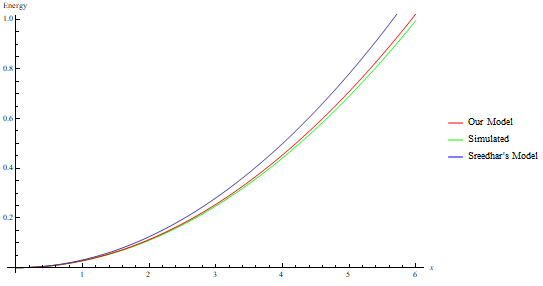}
\caption{(Left) The plot for $\alpha=1.5$. (Right) The plot for $\alpha=3.5$. In the left panel both results, ours (red curve) and \cite{sreedhar} (blue curve) agree with the simulated result (green curve) whereas in the right panel, for larger $\alpha$, the blue line \cite{sreedhar} deviates significantly from the green line but our result (red line) is more in agreement with the simulated result (green line).} \label{fig6}
    \label{fig:my_label}
\end{figure}

We obtained 6 consecutive eigenvalues for the system, and fitted the data, with a continuous function of the form,
\begin{equation}\label{fit}
y(x) = Ax^2 + Bx + C
\end{equation}

The results are plotted in Fig. \ref{fig6}.
The X-Axis contains the integers $n$, denoting the order of the energy eigenstate, and the Y-Axis contains the value of the energy.
The linear term in the fit, (\ref{fit}), is to accommodate any errors arising from the lowest eigenvalue obtained in this analysis, not being the ground-state.

We see from Fig. \ref{fig6}, how, with increasing $\alpha$, our model is able to match the simulated plot, while we notice no change in the plot(blue) depicting the model obtained by Sreedhar \cite{sreedhar}. This is clearly not a good approximation, since the energy eigenvalues does indeed change with $\alpha$, even in the thin-torus limit, and the effect is more pronounced for large $\alpha$ values, as discussed in Sub-section-\ref{subsec:results}.

\section{Conclusion and open problems}\label{sec:conclusion}
The present work deals with the quantum mechanics of a point particle moving along a torus knot on a regular torus. Novelty of our model and new results are the following.\\
(i) We have correctly taken into account the torsion and curvature terms in the Hamiltonian.\\
(ii) We have explicitly shown how the nature of  the  torus and torus knot influences magnitude of the torsion and curvature terms. This has allowed us to correctly consider those cases where the curvature effect is considerably larger than the torsion effect. Subsequently we have studied an approximate model where the torsion term is ignored.\\
(iii) We have computed energy eigenvalues and eigenfunctions in the thin torus approximation (with torsion neglected) both analytically as well as numerically. The new feature in our result is that, the energy depends on both $q,p$ that depicts the nature of the knot. We have also revealed an invariant structure, constructed out of the torus and torus knot properties.\\
(iv) In a series of previous works, one of us \cite{sg2,sg3}, have considered the Hannay angle and Berry phase for a similar problem with the torus itself revolving around its vertical axis. However  torsion and curvature effects were not considered in those works. In future we would like to pursue the study of Hannay angle and Berry phase for particle on a torus knot with torsion and curvature effects taken into account. We hope that the topologically non-trivial nature of the torus knot will induce new results.
It would also be interesting to study the motion of a particle with non-zero spin on the torus knot. Finally, the generalised system of a charged particle with spin, moving along the torus knot can be placed in an external electromagnetic field, to study the interaction effects. 

{\bf Acknowledgment: } DB  wishes to thank the Physics and Applied Mathematics Unit, Indian Statistical Institute, for a visit during which major part of the work was done.
\bibliographystyle{unsrt}  


\begin{thebibliography}{9}

\bibitem{schmidt} O. G. Schmidt and K. Eberl.
\textit{Thin solid films roll up into nanotubes}
Nature (London) 410, 168 (2001).

\bibitem{ahn}
J.H. Ahn et.al.
\textit{Heterogeneous three-dimensional electronics by use of printed semiconductor nanomaterials} 
, Science 314, 1754 (2006).

\bibitem{mei}
 Y. Mei et.al. \textit{High Performance Silicon Nanowire Field Effect Transistors}, Nano
Lett. 7, 1676 (2007).

\bibitem{ko}
 H. Ko et.al. \textit{Ultrathin compound semiconductor on insulator layers for high-performance nanoscale transistors},   Nature 468, 286 (2010).
 
\bibitem{park}
 S.I. Park, A.P. Le, J. Wu, Y. Huang, X. Li, and J. A. Rogers. 
\textit{Light emission characteristics and mechanics of foldable inorganic light-emitting diodes},
Adv. Mater. 22, 3062 (2010).

\bibitem{de}  B. S. DeWitt, Rev. Mod. Phys. 29, 377 (1957)

\bibitem{koppe} 
H. Jensen and H. Koppe.
\textit{Quantum mechanics with constraints}.
Annals of Physics, Volume 63, Issue 2, Pages 586-591, April 1971.

\bibitem{costa} 
R.C.T. da Costa.
\textit{ Quantum mechanics of a constrained particle}.
Phys. Rev. A, Vol-23, (1981) (1982).
\bibitem{Wang}
Yong-Long Wang, Meng-Yun Lai, Fan Wang, Hong-Shi Zong, and Yan-Feng Chen.
\textit{Geometric effects resulting from square and circular confinements for a particle
constrained to a space curve.}
Physical Review A. Vol-97, (042108), 2018.

\bibitem{ortix} 
C. Ortix
\textit{
Quantum mechanics of a spin-orbit coupled electron constrained to a space curve.}
PHYSICAL REVIEW B 91, 245412 (2015)


\bibitem{silva}
L.C.B da Silva, C.C. Bastos , and F.G. Ribeiro.
\textit{Quantum mechanics of a constrained particle and the problem of prescribed geometry-induced potential}.
Annals of Physics, Volume 379, Pages 13-33, April 2017.

\bibitem{pi} S. Matsutani.
, \textit{Path integral formulation of curved low dimensional space}, J. Phys. Soc. Japan 61 (1992) 3825;  \textit{Quantum field theory on curved low-dimensional space embedded in three-dimensional space}, Phys. Rev. A
47 (1993) 686.

\bibitem{ei} M. Encinosa, B. Etemadi., 
\textit{Energy shifts resulting from surface curvature of quantum nanostructures.}, Phys. Rev. A 58
(1998) 77.

\bibitem{grave} J. Gravesen, M. Willatzen. \textit{Eigenstates of Mobius nanostructures including curvature effects.}, Phys. Rev. A 72 (2005)
032108.

\bibitem{em} M. Ikegami, Y. Nagaoka.
\textit{, Electron motion on a curved interface}, Surface Science 263 (1992) 193

\bibitem{ferrari}G. Ferrari, G. Cuoghi.
\textit{Quantum mechanics on curved 2d systems with electric and magnetic fields}, Phys. Rev. Lett. 100
(2008) 230403.

\bibitem{oliv}G. de Oliveira.
\textit{Quantum dynamics of a particle constrained to lie on a surface}, J. Math. Phys. 55 (2014) 092106.

\bibitem{silva2} E. O. Silva, S. C. Ulhoa, F. M. Andrade, C. Filgueiras, R. G. G. Amorim.
\textit{Quantum motion of a point particle in the
presence of the Aharonov-Bohm potential in curved space}, Ann. Phys. 362 (2015) 739.

\bibitem{con} C. Filgueiras, F. Moraes.
\textit{On the quantum dynamics of a point particle in conical space}, Ann. Phys. 323 (2008) 3150.

\bibitem{filg} C. Filgueiras, E. O. Silva, F. M. Andrade.
\textit{Nonrelativistic quantum dynamics on a cone with and without a constraining
potential}, J. Math. Phys. 53 (2012) 122106.

\bibitem{Du} L. Du, Y.-L. Wang, G.-H. Liang, G.-Z. Kang, X.-J. Liu, H.-S. Zong. \textit{Curvature-induced bound states and coherent electron
transport on the surface of a truncated cone}, Physical E 76 (2016) 28.
\bibitem{entin} M. V. Entin, L. I. Magarill. \textit{Spin-orbit interaction of electrons on a curved surface}, Phys. Rev. B 64 (2001) 085330

\bibitem{gentile}P. Gentile, M. Cuoco, C. Ortix. \textit{Curvature-induced Rashba spin-orbit interaction in strain-driven nanostructures}, SPIN 3
(2013) 1340002.

\bibitem{santos} F. Santos, S. Fumeron, B. Berche, F. Moraes. \textit{Geometric effects in the electronic transport of deformed nanotubes},
Nanotechnology 27 (2016) 135302. 

\bibitem{marchi}
A. Marchi, S. Reggiani, M. Rudan, A. Bertoni. \textit{Coherent electron transport in bent cylindrical surfaces}, Phys. Rev. B 72
(2005) 035403.

\bibitem{krej} D. Krejcirık. \textit{Quantum strips on surfaces}, J. Geom. Phys. 45 (2003) 203.;

\bibitem{peter}J. Stockhofe and P. Schmelcher.
\textit{Nonadiabatic couplings and gauge-theoretical structure of curved quantum 
waveguides}, 
Physical Review A 89, 033630 (2014).
\bibitem{campo}
 A. del Campo, M. G. Boshier, A. Saxena. \textit{Bent waveguides for matter-waves: supersymmetric potentials and reflectionless
geometries}, Sci. Rep. 4 (2014) 5274.

\bibitem{haag} S. Haag, J. Lampart, S. Teufel, \textit{Generalised quantum waveguides}, Ann. Henri Poincare 16 (2015) 2535.

\bibitem{nanow} H.F. Zhang, C.M. Wang, and L.S. Wang. \textit{Helical Crystalline SiC, SiO2 Core−Shell Nanowires}, Nano. Lett. 2, 941
(2002).

\bibitem{nanoc} S. Xu et.al.
\textit{Assembly of micro/nanomaterials into
complex, three-dimensional architectures
by compressive buckling.}
Science 347, 154 (2015).

\bibitem{jo} V.F.R. Jones. \textit{A new polynomial invariant of knots and links}, Bull. Amer. Math. Soc. 12 103112 (1985). 

\bibitem{wit1} E. Witten. \textit{Quantum field theory and the Jones polynomial},
Comm. Math. Phys. 121 (1989), no. 3, 351399. 
\bibitem{wit2}E. Witten. \textit{Topological quantum field theory}, Comm. Math. Phys. 117 (1988), no.
3, 353386.  
\bibitem{tem}E. H. Lieb. \textit{Relations between the ‘percolation’ and ‘colouring’ problem and other graph-theoretical problems associated with regular planar lattices: some exact results for the ‘percolation’ problem}, Proc. Roy. Soc. London Ser. A 322 (1971), no. 1549, 251280.

\bibitem{fad} L. Faddeev and A. J. Niemi.
\textit{Knots and Particles},
Nature 387:58,1997 DOI: 10.1038/387058a0 (arXiv:hepth/9610193).

\bibitem{ke} H. Kedia, I. Bialynicki-Birula, D. Peralta-Salas, and W. T. M. Irvine. \textit{Tying Knots in Light Fields}, Phys. Rev. Lett. 111,
150404

\bibitem{new1}J. D'Ambroise, P. G. Kevrekidis and P. Schmelcher.    
\textit{Bright Solitary Waves on a Torus: Existence, Stability and Dynamics for 
the Nonlinear Schrödinger Model}, arXiv:1906.06001.

\bibitem{kh} F.C. Khanna, A.P.C. Malbouisson, J.M.C. Malbouisson, A.E. Santana.
\textit{Quantum field theory on toroidal topology: Algebraic structure and applications}, Physics Reports, vol.
539, p. 135-224, 2014 DOI: 10.1016/j.physrep.2014.02.002 arXiv:1409.1245.

\bibitem{oh} Y. Ohnuki and S. Kitakado, On \textit{Quantum Mechanics on a Compact Space}, Modern Physics
Letters A 7 (1992) 2477.

\bibitem{col} S. Coleman. \textit{Aspects of Symmetry}, Cambridge University Press, (1988).

\bibitem{raj} R. Rajaraman. \textit{Solitons and Instantons}, North-Holland Personal Library, (2003)

\bibitem{wil} F. Wilczek. \textit{Fractional Statistics and Anyon Superconductivity}, World Scientific Publishing
Company Pvt. Ltd. (1990).

\bibitem{sreedhar} 
V.V. Sreedhar.
\textit{The classical and quantum mechanics of a
particle on a knot}.
Annals of Physics, Volume 359, Pages 20-30, August 2015.

\bibitem{ohnuki} 
Y. Ohnuki and S. Kitakado. 
\textit{On Quantum Mechanics on Compact Space}. 
Modern Physics Letters A, Vol. 07, No. 27, pp. 2477-2482 (1992).

\bibitem{flo} R. Floreanini, R. Peracci, E. Sezgin. \textit{Quantum Mechanics on the circle and $W(1+\infty)$},
Phys.Lett. B271 (1991) 372-376.  

\bibitem{taka}S. Takagi and T. Tanzawa, \textit{Quantum Mechanics of a Particle Confined to a Twisted Ring}, Progress of Theoretical Physics, Vol. 87, No.3, March 1992.

\bibitem{moore} 
J.E. Moore
\textit{The birth of topological insulators}.
Nature, Vol-464, 11 March 2010.

\bibitem{sg}
P. Das, S. Pramanik, S.Ghosh. 
\textit{Particle on a Torus Knot:
Constrained Dynamics and Semi-Classical
Quantization in a Magnetic Field}.
Annals of Physics, Volume 374, p:67-83.(2016)

\bibitem{mathieu1} 
S.A. Wilkinson, N. Vogt, D.S. Golubev, J.H. Cole.
\textit{Approximate solutions to Mathieu's equation}.
\\\texttt{arXiv:1710.00657}(2017).

\bibitem{mathieu2} 
N.W. McLachlan.
\textit{Theory and Application of Mathieu Functions}
New York, Dover Publications (1964, 1947).

\bibitem{ince} 
E.L. Ince.
\textit{ On a general solution of Hills Equation}
Monthly Notices of the Royal Astronomical Society, Vol. 75, p.436-448,(1915).


\bibitem{sg2}	S. Ghosh. \textit{Particle on a Torus Knot: Anholonomy and Hannay Angle}, Int.J.Geom.Meth.Mod.Phys. 15 (2018) no.06, 1850097.
\bibitem{sg3}S. Ghosh. \textit{Geometric Phases for Classical and Quantum Dynamics: Hannay angle and Berry Phase for Loops on a Torus}, Int. J. Theo. Phys. (2019). https://doi.org/10.1007/s10773-019-04169-6, (arXiv:1905.03491).
\end{thebibliography}

\end{document}